\documentclass[twocolumn,showpacs,preprintnumbers,amsmath,amssymb]{revtex4}

\usepackage{graphicx}% Include figure files
\usepackage{dcolumn}% Align table columns on decimal point
\usepackage{bm}% bold math

\begin{document}
\preprint{APS}
\title{Vectorial atomic magnetometer based on coherent transients of laser absorption in Rb vapor}
\author{L. Lenci}
\author{A. Auyuanet}
\author{S. Barreiro}
\author{P. Valente}
\author{A. Lezama}
\author{H. Failache}

\email{heraclio@fing.edu.uy} \affiliation{Instituto de F\'{\i}sica, Facultad de Ingenier\'{\i}a, Universidad de
la Rep\'{u}blica,\\ J. Herrera y Reissig 565, 11300 Montevideo, Uruguay}
\date{\today}

\begin{abstract}
We have designed and tested an atomic vectorial magnetometer based on the analysis of the coherent oscillatory transients in the transmission of resonant laser light through a Rb vapor cell. We show that the oscillation amplitudes at the Larmor frequency and its first harmonic are related through a simple formula to the angles determining the orientation of the magnetic field vector. The magnetometer was successfully applied to the measurement of the ambient magnetic field.\\
\end{abstract}

\pacs{42.50.Gy, 07.55.Ge, 42.50.Md, 32.30.Dx}

\maketitle
\section{\label{Introduction}Introduction}

Most atomic magnetometers measure the modulus of a magnetic field by measuring directly or indirectly the Larmor frequency of the atomic magnetic moment in the presence of the external field \cite{Budker:2002,Alexandrov:2005,Budker:2007,Budker:2013}. For many applications as, for instance, geophysical measurements, it is  also important to determine the direction of the magnetic field.
Different methods have been proposed and realized to measure the magnetic field vector. A vectorial atomic magnetometers based on electromagnetically induced transparency (EIT) in sodium transition was proposed by Lee et al \cite{Lee:1998}. They have shown that the dependence of the phase shift between pump and probe fields on the angle between the magnetic
 field and the light propagation and polarization directions allows the measurement of both the magnitude and the direction of the magnetic field. Weis et al.\cite{Weis:2006}
 presented a theoretical study showing how the direction of the magnetic field vector can be extracted by the analysis of the spectra of an optical radio frequency double resonance
 magnetometer. Pustelny et al. \cite{Pustelny:2006} have studied experimentally and theoretically  a vectorial magnetometer based on nonlinear magneto-optical
 rotation (NMOR). The relative amplitude of the NMOR resonances in the $^{85}$Rb D1 line allowed the determination of the magnetic field direction. The dependence of the EIT resonances amplitudes on the direction of the magnetic field has also been studied by Yudin et al.
 \cite{Yudin:2010} and Cox et al. \cite{Cox:2011}. In others atomic magnetometers, the three components of the magnetic field
 vector were measured using Helmholtz coils to generate additional small magnetic fields \cite{Seltzer:2004,Alexandrov:2004}. These magnetometers based on the detection of coherent
 effects in the frequency domain can be used for the measurement of magnetic fields of the order of those usually found in geophysics.\\

\begin{figure}[h]
\includegraphics[width=7cm]{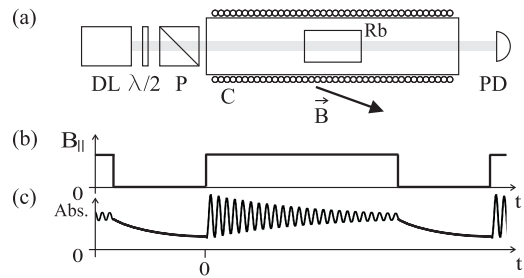}
\caption{(a) Scheme of the magnetometer setup, where DL: diode laser, P: polarizer, $\lambda/2$: half-wave plate, PD: photodetector, C: coil; (b) Temporal sequence of the total magnetic field component $B_{\|}$ along the light propagation direction and (c) corresponding laser absorption transient.} \label{figsetup}
\end{figure}

Recently several scalar atomic magnetometers were proposed based on the time domain analysis of coherence transients \cite{Lenci:2012,Breschi:2013,Behbood:2013}. In the present work we report the realization of a vectorial atomic magnetometer based on the time domain analysis of the coherent transient evolution of a Rb vapor sample probed with resonant laser light. Information about the magnetic field direction is extracted from the relative amplitudes of the transient oscillation at the atomic ground state Larmor frequency and its first harmonic.\\

\maketitle
\section{\label{Experimental setup}Magnetometer setup}

The proposed magnetometer setup (Fig.\ref{figsetup}) is similar to the one used for our previous scalar magnetometer \cite{Lenci:2012}. The principle of operation of the magnetometer is based on the use of
a sequence of two consecutive atom-light interaction intervals. During the first interval the total magnetic field component along the light propagation direction is canceled through the application of an additional magnetic field produced with a solenoid. During this interval the atomic sample becomes
aligned by optical pumping.

In a second interval, the applied field is turned off and the previously prepared atomic alignment evolves in the presence of the magnetic field to be measured. This evolution can be understood as a precession of the atomic alignment \cite{Budker:2001} around the magnetic field vector. The same linearly polarized laser beam, resonant with the atomic transition, is used for both
optical pumping and probing the transient evolution of the atomic ground state coherence. The transmitted light intensity is detected with a photodiode. The coherent atomic evolution results in oscillations of the light transmission at the ground state Larmor frequency $\omega_L$ and the first harmonic $2\omega_L $. The Larmor frequency is related to the magnetic field modulus $B$ through the expression: $\omega_L=g_{F}\mu_{B} B$ where $\mu_{B}$ is the Bohr magneton and $g_{F}$ is the ground-state hyperfine level Land\'e factor which is known with high accuracy \cite{Steck:2010}.

The relative amplitude of the transient oscillations at $\omega_L$ and $2\omega_L$ depends on the magnetic field vector direction and can be used to measure the magnetic field vector
components as it is theoretically analyzed in the next section.

\maketitle
\section{\label{Theory}Theory}

In this section, we present a simplified theoretical treatment based on a model transition from a ground level
with total angular momentum $F=1$ to an excited level with $F'=0$ (see inset in Fig. \ref{figtheory}). We use the notation $|F,m\rangle$ for the involved states where $F$ is the atomic level angular momentum and $m$ the magnetic quantum number. As discussed below, the results obtained can also be applied to other transitions.

The total Hamiltonian of the system is
$H=H_A + H_M + V_D$, where $H_A=\hbar\omega_o |0,0\rangle\langle0,0|$ is the atomic Hamiltonian and
$\hbar\omega_o$ is the energy difference between the excited and ground levels, $H_M= g_{F}\mu_{B}
\mathbf{B}.\mathbf{F}$ is the magnetic Hamiltonian where $\mathbf{B}$ is the magnetic field and $\mathbf{F}$ is the total angular momentum
operator. The atom-light interaction  is described by the term $V_D = -\mathbf{E}.\mathbf{D}$ where $\mathbf{E}$ is the light electric field and $\mathbf{D}$ is the dipole moment
operator.\\

We assume that the system evolves due to the presence of a magnetic field and consider the interaction with the light field as a perturbation.
 The density matrix $\rho(t)$ that describes the atomic state is given by $ \rho(t) = U(t) \rho_0 U^\dagger(t) $, where $\rho_0 $
is the atomic state at $t=0$ and $U(t)=e^{-i(H_A+H_M)t/\hbar}$ is the evolution operator.

\maketitle \subsection{Atomic alignment preparation}

We consider a laser beam propagating along the $z$ axis and polarized along the $x$ (see Fig.\ref{figtheory}a). During the preparation interval the total magnetic field is canceled (through the application of an auxiliary field) and the atomic system
is optically pumped to dark states. Using $x$ as the quantization axis, the light field only couples the states $|1,0\rangle$ and $|0,0\rangle$. In consequence the system is pumped into the initial state given by the density matrix $\rho_0=\frac{1}{2}\left(|1,1\rangle \langle 1,1|+|1,-1\rangle \langle 1,-1|\right)$ representing the alignment of the system.

\maketitle \subsection{Alignment precession}

\begin{figure}[h]
\includegraphics[width=8cm]{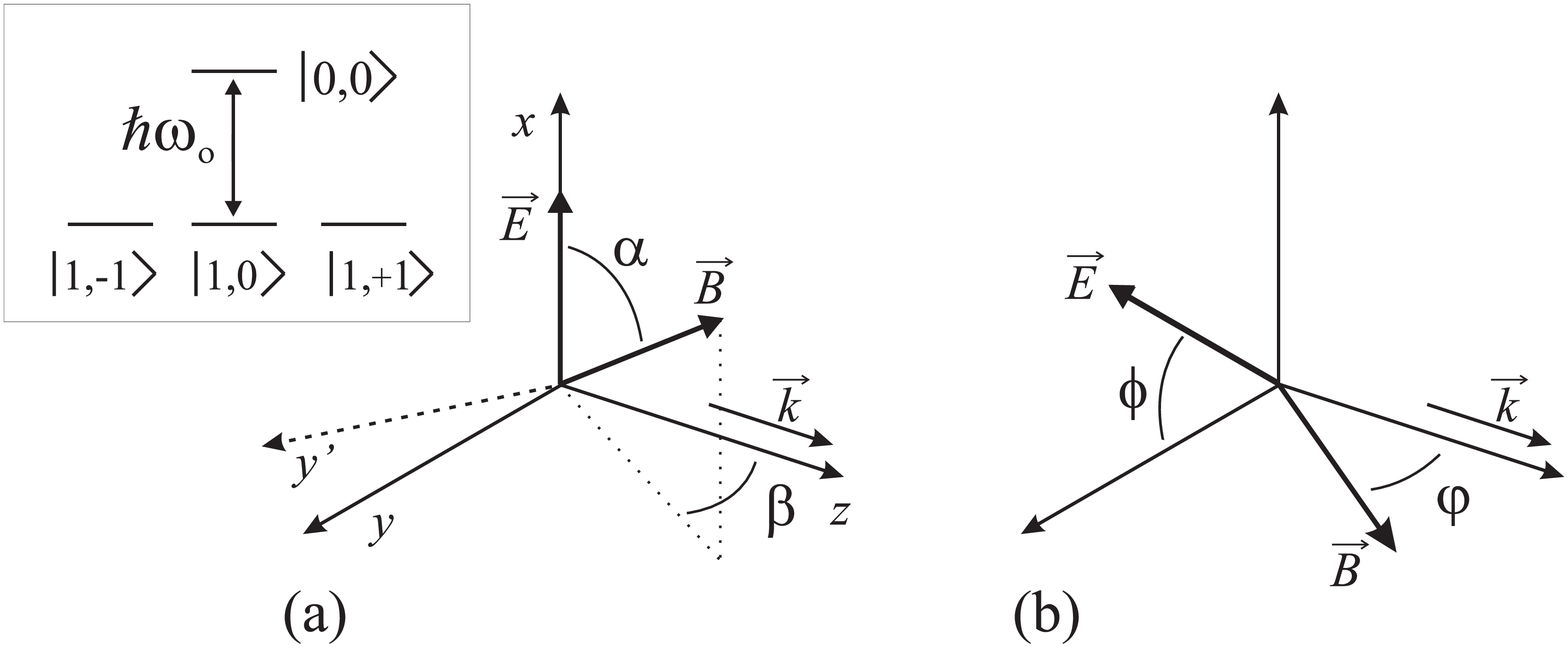}
\caption{(a) Light and magnetic field vectors. Inset: model atomic transition $F=1 \rightarrow F'=0$ using $x$ as the quantization axis. (b)
Alternative set of coordinates used for the experiment in which the magnetic field and the wave vector are perpendicular to one of the coordinate axis.} \label{figtheory}
\end{figure}

After the preparation interval, the auxiliary magnetic field is turned off and the atoms evolve in the presence of the magnetic field to be measured (Fig.\ref{figtheory}a). To simplify the calculation of the atomic system evolution it is convenient to describe the state of the system using a reference frame where the magnetic field direction corresponds to the quantization axis. This is achieved by two consecutive rotations around the $x$ and $y'$ axis ($y^{\prime}$ is obtained after rotation of the $y$ axis an angle $\beta$ around axis $x$. See Fig.\ref{figtheory}a). In the rotated frame, the initial density matrix is $\rho_{R}=R \rho_0 R^\dagger$
where $R=R_{y^{\prime}}(\alpha).R_x(-\beta)$ is the product of the rotation operators $e^{-i\beta \widehat{F}_x/\hbar}$ and $e^{i\alpha \widehat{F}_{y^{\prime}}/\hbar}$. The evolution of the system in this frame due to the magnetic field is described by $\rho(t)=U(t) \rho_R U^\dagger (t)$.

Finally, after turning back to the original frame, we compute the time dependence of the light absorption using standard perturbation theory to calculate the transition probability rate $W$
from the ground to the excited level: $W=\frac{2\pi}{\hbar}\langle 0,0|V\rho (t)V^{\dagger} |0,0\rangle$. We obtain:

\begin{multline}
      W \propto \sin^2 (\alpha) \Big[ \frac{5}{4}+\frac{3}{4} \cos (2\alpha)-\cos (\omega_L t)\Big( 1+\cos (2\alpha)\Big)\\
      -\frac{\cos (2 \omega_L t)}{4}\Big( 1-\cos (2\alpha)\Big) \Big]
      \label{eq5}
    \end{multline}

Two oscillating terms appear in the evolution; one is oscillating at the Larmor frequency $\omega_L$ and the other at its first harmonic $2\omega_L$. The ratio of the amplitude of the two oscillating terms directly relates to the angle $\alpha$:

\begin{equation}\label{eq6}
    b_{\omega_L}/b_{2\omega_L} = 4 \tan^{-2} \alpha
\end{equation}

The proposed magnetometer is based on the use of Eq.\ref{eq6} for the determination of the angle between the magnetic field vector and the light polarization direction. In the derivation of  Eq.\ref{eq5} we have ignored the system relaxation due to ground state decoherence. It is assumed that the decoherence characteristic time is much longer than the
Larmor period. In consequence, the magnetometer operation is limited to magnetic fields that are large enough to verify this assumption.\\

As expected, Eq.\ref{eq5} predicts that the transient response is zero for a magnetic field parallel to the light polarization direction ($\alpha=0$) since no precession occurs for an atomic
alignment that is created parallel to the magnetic field. Another singular configuration corresponds to the magnetic field perpendicular to the light polarization ($\alpha=\pi/2$), in which case only
the precession at $2 \omega_L$ is observed without the component at $\omega_L$ in consistency with Eq.\ref{eq6}.\\

In order to apply Eq.\ref{eq6} to actual experiments it is convenient to use the alternative angular coordinates shown at Fig.\ref{figtheory}.b best suited to experimental control  ($\phi$ measures the light polarization angle with the plane containing the
magnetic field and the light wave vector, $\varphi$ is the angle between the magnetic field and the light wave vector). Eq.\ref{eq6} can then be written as:

\begin{equation}\label{eq7}
   \frac{b_{\omega_L}}{b_{2\omega_L}}=\frac{4\cos^{2}{\phi}}{{\cot^{2}{\varphi}+\sin^{2}{\phi}}}
\end{equation}
\\
The derivation of Eqs.\ref{eq6} and \ref{eq7} was done under the assumption that the initial state
is a statistical mixture of coherent dark states $|1,+1\rangle$ and $|1,-1\rangle$  with no population in the $|1,0\rangle$ state. In order to ensure the preparation of this initial atomic
state, a measurement procedure is described in the next section.

Strictly speaking, Eqs.\ref{eq6} and \ref{eq7} were only derived for a $F=1 \rightarrow F'=0$ transition. In the next section we describe a numerical simulation showing that it can also be applied to different transitions including the $F=2 \rightarrow F'=1$ transition used in the experiment.\\

\maketitle \subsection{\label{measproc}Measurement procedure}

The initially prepared atomic state is in principle dependent on the magnetic field. Consequently, a specific procedure needs to be followed to ensure that the prepared state corresponds to the sate $\rho_0$ assumed in the theory.\\

First, the light polarization must be rotated to a direction perpendicular to the
magnetic field ($\phi=\pi/2$). In practice this can be done by rotating the light polarization to the direction that cancels the transient oscillation at the Larmor frequency $\omega_L$, only
preserving the transient oscillation at $2\omega_L$. Some information about the magnetic field direction is thus obtained since the plane containing the magnetic field vector and the light wave vector is identified. Also, the modulus of the magnetic field can be measured using the procedure described in \cite{Lenci:2012}. At this stage, a
perfect cancelation of the magnetic field during the state preparation interval via optical pumping is not essential. As discussed in \cite{Lenci:2012}, is sufficient to roughly cancel the magnetic field component along the light propagation to observe a large enough amplitude of the atomic signal.\\

Next, the light polarization is rotated an angle $\pi/2$ to the plane containing $\mathbf{B}$ and $\mathbf{k}$. In this stage the magnetic field
component along the light propagation direction $B_{\|}$ must be canceled during the preparation interval using the external coil. To optimize this field cancelation the maximization of the transient signal amplitude is used as a criterium. Once this cancelation is achieved, the total magnetic field during the preparation interval is parallel to the light polarization. In consequence, the laser light only induces $\pi$ transitions (taking the light polarization as quantization axis, see Fig. \ref{figtheory}) and the atomic system is pumped to a statistical mixture of the states $|1,+1\rangle$ and $|1,-1\rangle$ as was assumed in the derivation of Eq.\ref{eq6}. After the preparation interval, the current in the coil is turned off and the oscillatory transient observed. The magnetic field direction given by $\varphi$ is then determined with the help of Eq.\ref{eq7} using $\phi=0$.\\

Equation \ref{eq7} does not allow the determination of the sign of the angle $\varphi$. The orientation of the field component $B_{\bot}$ perpendicular to the direction $\mathbf{k}$ is not determined by this equation. However, the orientation of the
field component $B_{\|}$ is known since it is  determined by the sign of the external field required to compensate such component during the preparation interval. If a priori knowledge of the sign of $\varphi$ is not available, an additional measurement in the presence of an additional magnetic field perpendicular to light propagation (produced by another coil) can be used to determine this sign.\\

\begin{figure}[h]
\includegraphics[width=8cm]{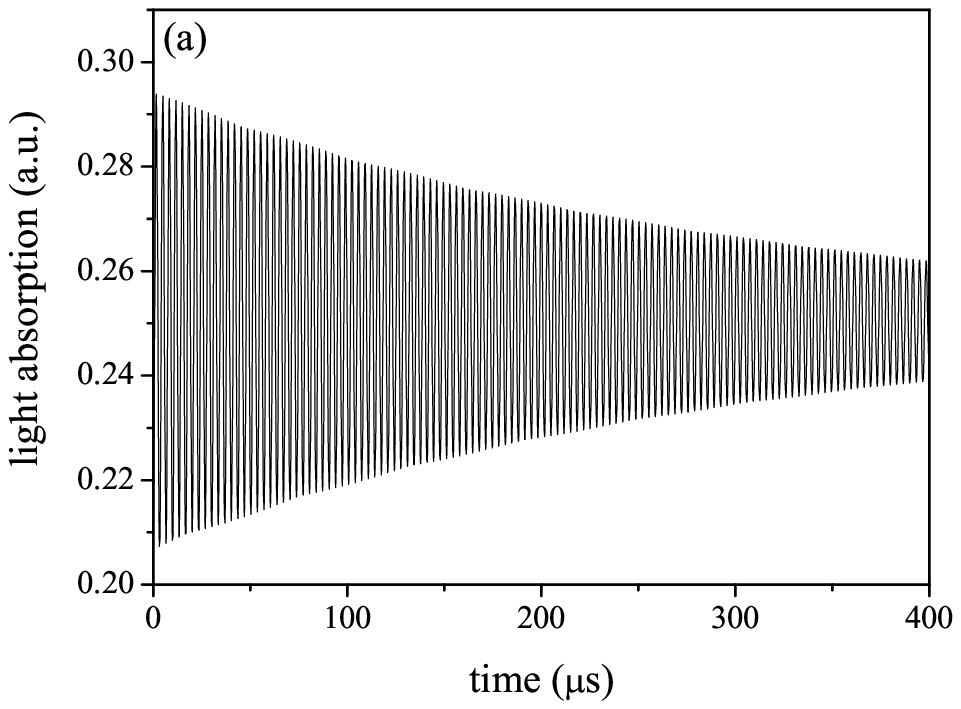}
\includegraphics[width=8cm]{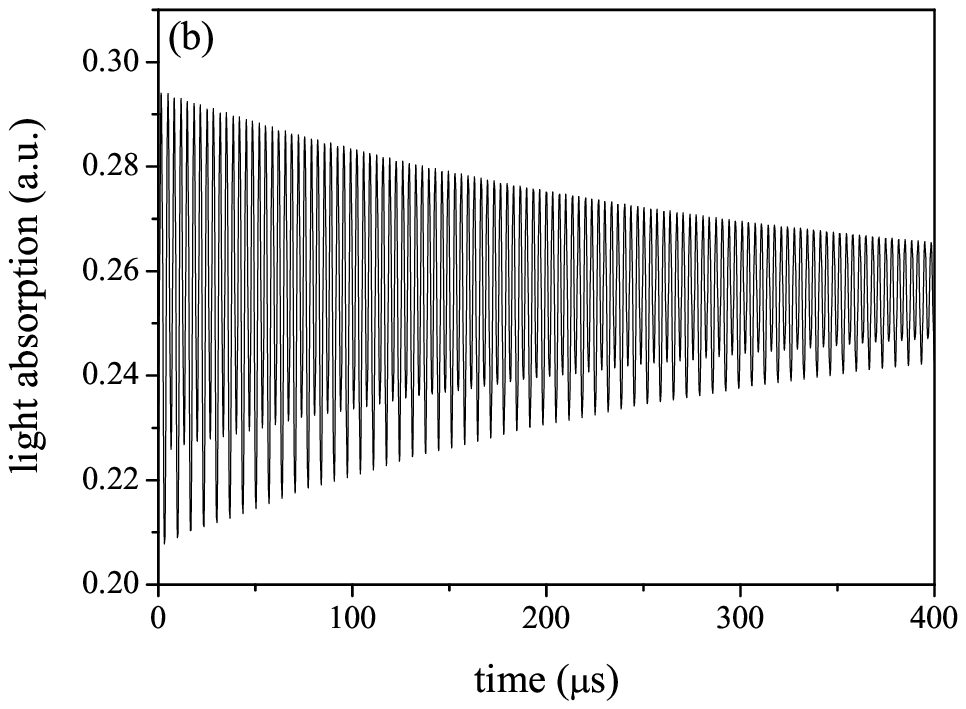}
\caption{Numerical prediction of transients for (a) $\phi=\pi/2$, (b) $\phi=0$.} \label{fig5}
\end{figure}

The imperfect cancelation of the magnetic field component $B_{\|}$ during the preparation interval introduces a measurement error. It was estimated using the numerical simulation
described in the next section.\\

\begin{figure}[h]
\includegraphics[width=8cm]{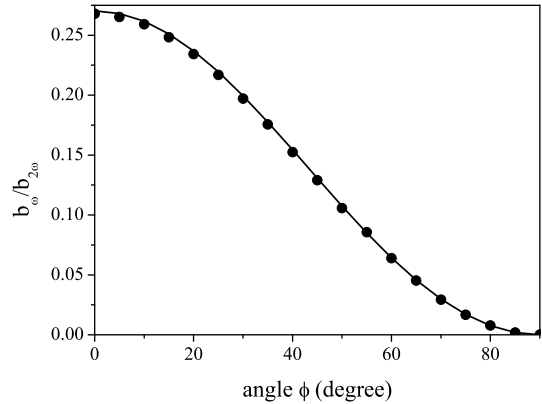}
\caption{Comparison of ratio $b_{\omega}/b_{2\omega}$ determined by fitting the numerical simulation for a transition $(F=2 \rightarrow F'=1)$ (dots) and the prediction after the expression Eq.\ref{eq6} (line).} \label{fig6}
\end{figure}

\maketitle
\section{\label{Numerical}Numerical simulations}

We have computed theoretical plots of the transient evolution of the laser beam absorption by Rb vapor by numerically solving  the optical Bloch equations including all Zeeman sub-levels \cite{Valente:2002}. Typical results are shown at Fig.\ref{fig5}(a) and
\ref{fig5}(b) for $\phi=\pi/2$ and $\phi=0$ respectively for the $F=2 \rightarrow F'=1$ transition used in the experiment. The parameters used in the numerical model are determined from the experimental conditions.\\

The calculated transients are very well adjusted to the damped oscillation function \cite{Valente:2002}:

\begin{align}\label{eq9}
A(t)=e^{-\gamma_{1}t}(b_{2\omega_L}\cos({2\omega_L t + \theta_{2\omega_L}})+
\nonumber \\
b_{\omega_L}\cos({\omega_L t + \theta_{\omega_L}}))+ C+De^{-\gamma_{2}t}
\end{align}

\begin{figure}[h]
\includegraphics[width=8cm]{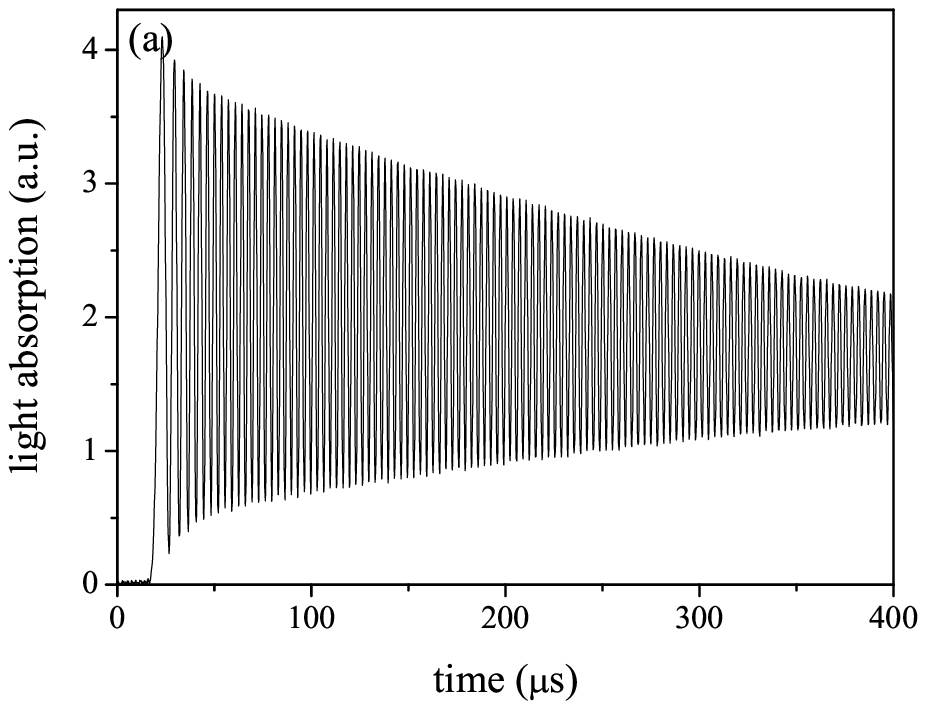}
\includegraphics[width=8cm]{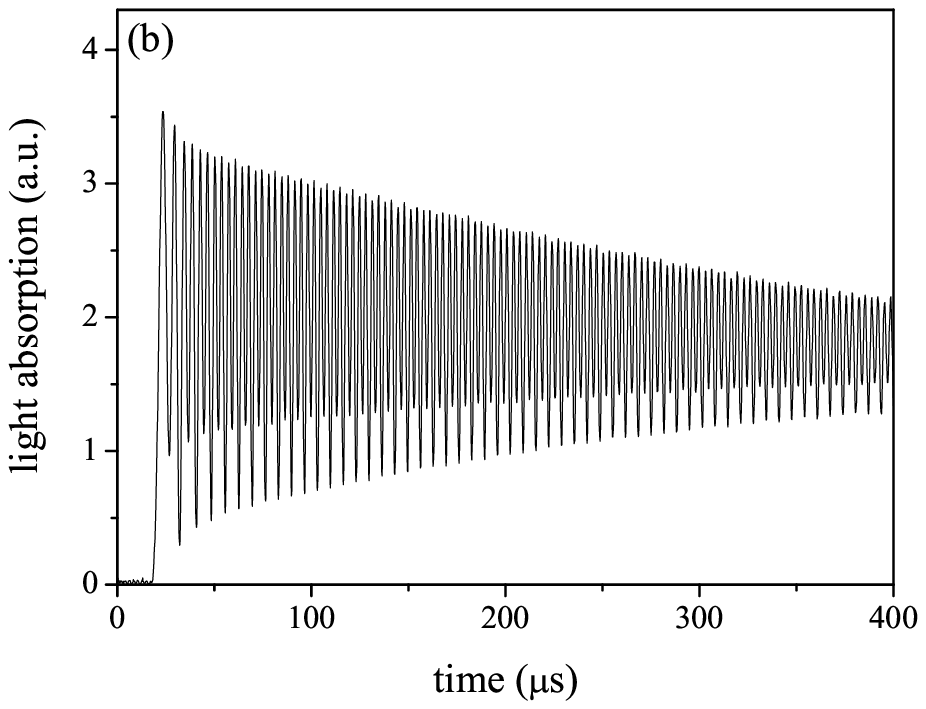}
\caption{Experimental absorption transients measured for the atomic cell in the environmental magnetic field for a) $\phi=\pi/2$, and b) $\phi=0$.} \label{fig4}
\end{figure}

After fitting the calculated transients to Eq.\ref{eq9} we determine the ratio $b_{\omega_L}/b_{2\omega_L}$. The result for the $^{87}$Rb, $F=2 \rightarrow
F'=1$ transition  used in the experiments is shown in Fig.\ref{fig6}. Also shown in this figure is the prediction from Eq.\ref{eq7}. The comparison of these plots shows that Eq.\ref{eq7} is acceptably accurate in spite of having been derived for a different $(F=1 \rightarrow F'=0)$ transition. The approximate validity of  Eq.\ref{eq7} was numerically checked for all the Rb D transitions.\\

The numerical model was also used to estimate the magnetometer uncertainty. As mentioned in Sec. \ref{measproc}, an error is introduced if the magnetic field along the light propagation direction is not properly compensated
during the atomic state preparation interval. This compensation is achieved through the maximization of the oscillatory transient amplitude. Assuming that the maximization is
done with an uncertainty of $1\%$ then the uncertainty on $\varphi$ is estimated to be about  $0.1 ^{\circ}$ in the conditions of the experiment.

\maketitle
\section{\label{Experiment}Experiment}

We have used a CW diode laser tuned to resonance with the $^{87}$Rb, $F = 2 \rightarrow F =1$ transition of the D1 line ($795 nm$). An auxiliary Rb cell was used to stabilize
the laser frequency on the Doppler absorption profile. The laser beam was expanded and an $8 mm$ diaphragm selected the center of the beam to obtain an intensity homogeneity better
than $10\%$ cent. Neutral density filters were used to obtain $25 \mu W$ radiation power at the atomic cell. The polarization was controlled with a linear polarizer. A half
wave plate was used to match de diode laser polarization to the polarizer. The $5 cm$ long Rb glass cell has $2.5\ cm$ diameter windows. A silicone tube with circulating
hot water was wrapped around the cell to heat it to $55^{o}C$ without introducing a stray magnetic field. The cell contains both $Rb$ isotopes in
natural abundance and $30 Torr$ of $Ne$ as a buffer gas.\\

In a first experiment a magnetic field produced under controlled conditions was measured placing the cell inside a $15 cm$ diameter, $40 cm$ long solenoid whose axis formed an angle of $\varphi \sim 14 ^{o}$
with the light beam propagation vector. The whole system was inserted in a three layers mu-metal shield. \\

During the alignment preparation time interval, the total magnetic field is canceled by switching off the electric current in the solenoid. When the magnetic field is turned on, the transient damped oscillation on the transmitted laser light intensity is measured as a function of the light polarization angle $\phi$. The spectra measured for $\phi=\pi/2$ and $ \phi=0$ are well reproduced by the numerically simulated transients shown at Fig.\ref{fig5}, however some differences arise due to the magnetic
field inhomogeneity existing in the cell. As the dimension of the laser beam is not negligible respect to the solenoid diameter, a transversal magnetic field gradient is
present in the atom-light interaction region introducing a slight spread of the Larmor frequency that modifies the envelope of the experimental oscillatory transients.

After numerical fitting of the experimental measurements we found that $\varphi = 13.6 ^{\circ}$. Due to the magnetic field spatial inhomogeneity it is difficult to evaluate the uncertainty of this measurement.\\

The magnetometer operation was also tested placing the cell outside the $\mu$-metal shield to measure the ambient magnetic field. The magnetometer was placed in an empty room to reduce the influence of inhomogeneous and fluctuating magnetic fields usually present in the laboratory.

The measurement was done following the procedure described in Sec. \ref{measproc}. First the polarizer was rotated until only one frequency was observed in the oscillatory transient
as shown at Fig.\ref{fig4}.a. The polarization transmission direction determines the direction of the magnetic field component perpendicular to $\mathbf{k}$. The polarizer was then rotated by $\pi/2$ and the observed transient used to determine the angle $\varphi$ via Eq.\ref{eq7} (see Fig.\ref{fig4}.b). The measured modulus of the magnetic field was $B = 22491 \pm 1 nT$ with a  direction given by $\varphi = 21.2 \pm 0.1 ^{\circ}$ \cite{note1}.\\

\begin{figure}[h]
\includegraphics[width=8cm]{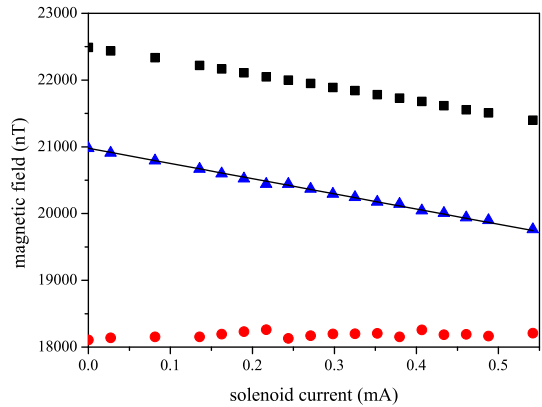}
\caption{(Color on-line) Experimental measurement of the environmental magnetic field vector components $B_{\bot}$ (red circles) and $B_{\|}$ (blue triangles) when an additional magnetic field along $B_{\|}$ is added [A constant value of $10^4 nT$ was added to $B_{\bot}$ for clarity]. Magnetic field modulus (black squares). Estimated value of $B_{\|}$ deduced from coil calibration (solid line)} \label{fig7}
\end{figure}

A series of measurements was performed introducing with the solenoid a well known magnetic field along the light propagation direction. It was checked that the magnetometer properly describes the magnetic field vector variation as it was systematically modified. As shown in Fig.\ref{fig7}, the field components along $\mathbf{k}$ is gradually reduced by the  field introduced by the solenoid while the component
perpendicular to $\mathbf{k}$ is consistently not modified. Also shown in Fig. \ref{fig7} are the measured magnetic field modulus and the variation of $B_{\|}$ deduced from an independent calibration of the solenoid.\\

\maketitle
\section{Conclusion}

We have proposed a vectorial atomic magnetometer based on the time domain measurement of the atomic absorption oscillatory transients induced by the atomic alignment
precession around the magnetic field. A simple formula relates the magnetic filed direction with the amplitude of these oscillatory transients. The suggested magnetometer is well adapted to the measurement of slow varying fields such as the Earth's magnetic field.\\

\maketitle
\section{\label{Acknoledgments}Acknoledgments}
We wish to thank A. Saez for its help with the experiment setup. This work was supported by CSIC, ANII and PEDECIBA (Uruguayan agencies).\\

\bibliographystyle{apsrev}
%\bibliography{References}

\begin{thebibliography}{18}
\bibitem[{not()}]{note1}
\bibinfo{note}{The relatively small value of the ambient magnetic field is due
  to the fact that Uruguay is situated at the South Atlantic Magnetic Anomaly.}
\end{thebibliography}

\end{document}